\documentclass[twocolumn]{revtex4}

\usepackage{graphicx}

\begin{document}

\title{
Plasma mechanisms of resonant terahertz detection  in
two-dimensional electron channel with split gates 
}

\author{ 
V.~Ryzhii\footnote{Electronic mail: v-ryzhii@u-aizu.ac.jp}
and A.~Satou
}
\address{
Computer Solid State Physics Laboratory, University of Aizu, 
Aizu-Wakamatsu 965-8580, Japan} 
\author{T. ~Otsuji}  
\address{Research Institute of Electrical Communication,
Tohoku University, Sendai 980-8577, Japan
}
\author{M.~S.~Shur}  
\address{Department of Electrical, Computer, and Systems Engineering, 
 Rensselaer Polytechnic Institute, Troy, NY 12180
}

\begin{abstract}
We 
analyze the operation of
a resonant detector of
terahertz (THz)
radiation based on a two-dimensional electron gas (2DEG) channel
with split gates. 
The side gates are used for the excitation of plasma oscillations
by incoming THz radiation and control of the resonant plasma frequencies.
The central gate provides the potential barrier separating
the source and drain portions of the 2DEG channel.
Two possible mechanisms of the detection
are considered: (1) modulation of the ac potential drop
across the barrier and (2) heating of the 2DEG due to the resonant 
plasma-assisted
absorption of  THz radiation followed by an increase in thermionic 
dc current through the barrier. 
Using the device model
we calculate  the frequency and temperature
dependences of the detector responsivity associated with both
dynamic and heating (bolometric) mechanisms. It is shown that 
the dynamic mechanisms dominates at elevated temperatures,
whereas the heating mechanism provides larger contribution at low
temperatures, $T  \lesssim 35 - 40$~K.
\end{abstract}

\maketitle

\newpage
\section*{I.~INTRODUCTION}

Plasma oscillations in heterostructures with a two-dimensional electron gas (2DEG)
and some devices using the excitation of these oscillations
have been experimentally  studied over decades
(see, for instance, 
Refs.~\cite{1,2,3,4,5,6,7,8,9,10,11,12,13,14,15,16,17}). 
There are also many theoretical
papers on different aspects of these plasma waves and their potential
applications.
The gated and ungated 2DEG channels in heterostructures can serve 
as resonant cavities for terahertz (THz) electron 
plasma waves.~\cite{18}
The resonant properties of such channels can be utilized in
different THz devices, in particular, resonant 
detectors.~\cite{3,4,5,9,10,12,13,14,15,16,17}
The mechanism of the THz detection observed in transistor
structures~\cite{3,4,6,7,9,10,15,16,17}
might be attributed to the nonlinearity of the plasma oscillations
as suggested previously.~\cite{18}
The variation of the conductivity of 2DEG with periodic gate
system can be possibly explained by the heating of
2DEG by absorbed THz radiation 
(heating or bolometric mechanism).~\cite{5,13}
Recently, a concept of resonant detectors in which 
a plasma resonant cavity
is integrated with a Schottky junction has been proposed 
and substantiated.~\cite{19,20}
As shown, when the frequency of incoming THz radiation
is close to one of the plasma resonant frequencies,
the ac potential drop across the barrier at
the Schottky junction becomes resonantly
large. This leads to rather large values of the rectified component
of the current through the junction, which is used as output signal
of the detector.
Since the resonant excitation of the plasma oscillations by
absorbed THz radiation leads to the heating of 2DEG,
the heating (bolometric) mechanism can also contribute
to the rectified current over the Schottky barrier.
Thus, two mechanisms can be responsible for the
operation of this detector: the dynamic mechanism considered
previously~\cite{19} and the heating mechanism.
Similar mechanisms can work in the detectors utilizing
the excitation of plasma oscillations but having the barrier
of another origin.  In particular, the barrier formed 
by an additional
gate which results in an essential depletion under
it was used in recent experiments.~\cite{14}
In this paper, we develop a device model for
a THz resonant detector based on a heterostructure with the 2DEG channel
and the barrier region utilizing the plasma oscillations excitation.
We consider a device with three gates (see Fig.~1):
the 2DEG under two extreme gates forms the plasma resonant cavities,
while the central gate, to which a sufficiently 
large  negative bias voltage  is applied, forms the barrier.
The excitation of the plasma oscillations by the incident THz
radiation is assumed to be associated with an antenna
connected with the source and drain contacts to the 2DEG channel.
In the following we demonstrate that each of the mechanisms
in question can dominate in different temperature ranges
(the dynamic mechanism is predominant at elevated temperatures,
whereas the heating mechanism provides larger contribution at low
temperatures, $T  \lesssim 35 - 40$~K).
The device structure under consideration differs from that
studied in Ref.~\cite{14} in which the THz radiation input
was realized by a periodic grating. However,
the physical mechanisms of detection are the same.

\section*{II.~EQUATIONS OF THE MODEL}

We assume that the net drain-source voltage $V$ includes both the dc bias voltage
$V_d$ and the ac component $V_{\omega}\cos\omega t$ induced by incoming 
Thz radiation via an antenna connected with 
the source and drain contacts.
Thus, $V = V_d + V_{\omega}\cos\omega t$, where $\omega$ is the THz radiation
frequency. The bias voltage $V_g$ is applied between the side gates
and the pertinent contacts. This voltage affects the electron density
in the quasi-neutral portions of the 2DEG channel, so that
the dc electron density is given by
\begin{equation}\label{eq1}
\Sigma_0 = \Sigma_d + \frac{\ae V_g}{4\pi e W},
\end{equation}
where $e$ is the electron charge, $W$ is the gate layer thickness, and
$\ae$ is the dielectric constant.
It is also taken to be that the  potential drop, $V_{cg}$,
between the central gate and the source contact is negative 
and its absolute value is sufficiently large 
to deplete some
portion of the 2DEG channel underneath this gate and create
a potential barrier. This implies that
$V_{cg} <  - 4\pi e W/\ae\Sigma_d = V_{th}$, where
$V_{th}$ is the 2DEG threshold voltage.

\begin{figure}[b]
  \begin{center}
    \includegraphics[width=8.5cm]{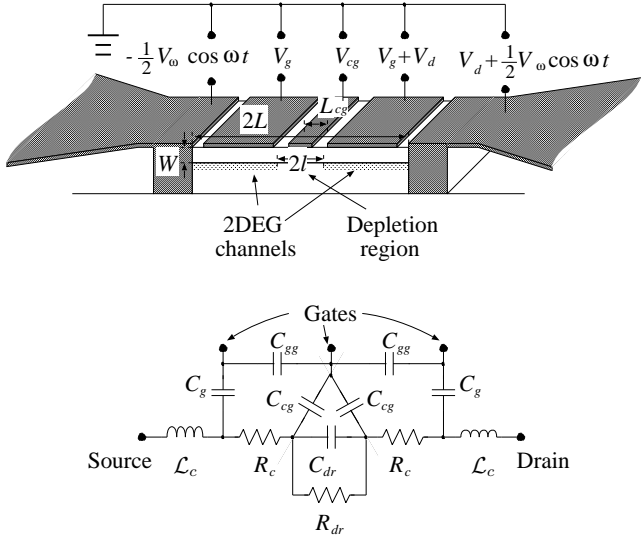}
    \caption{Schematic view of the THz detector under consideration
      and its simplified equivalent circuit.}
  \end{center}
\end{figure}

The ac component of the electric potential in the channel
$\delta\varphi$ satisfies the boundary conditions which follow
from the fact that at the source and drain contacts its values
coincide with the ac components of the incoming signal voltage:
\begin{equation}\label{eq2}
- \delta\varphi\,|_{x = - L} =  \delta\varphi\,|_{x = + L} 
= \frac{1}{2}V_{\omega}\cos\omega t.
\end{equation}
Assuming that at the conductivity
of the quasi-neutral sections is much larger  than that of
depletion region (including the thermionic current over the barrier
and the displacement current associated with the capacitances
shown in the equivalent circuit of Fig.~1)
  one may use the following
additional conditions:
\begin{equation}\label{eq3}
\frac{\partial}{\partial x}\delta\varphi\biggl|_{x = \pm l}
\simeq 0.
\end{equation}
Here $2L$ is the length of the 
2DEG channel between the source and drain contacts,
$2l > L_{cg}$ is the length of the depletion region under 
the central gate, and $L_{cg}$ is the length of the latter (see Fig.~1).
The quantity $l$ is determined by the potential of the central gate 
$V_{cg}$. 
Definitely, the  conductivity of the quasi-neutral sections
markedly exceeds the real part of the depleted region admittance. 
The role of the displacement current
across the depleted region is discussed in VIII.

The electron  transport in the channel is governed by the hydrodynamic
equations coupled with the Poisson equation for the self-consistent electric potential~\cite{18}. At sufficiently small intensities of THz radiation,
this system of equations can be linearized and reduced to
the following equation~\cite{21,22} valid in the quasi-neutral regions of the channel
($- L \leq x < -l$ and $l < x \leq L$): 

\begin{equation}\label{eq4}
\frac{\partial}{\partial t}\biggl(\frac{\partial}{\partial t} + 
\nu\biggr)
\delta \varphi = s^2\frac{\partial^2}{\partial x^2}\delta \varphi,
\end{equation}
where $\nu$ is the  frequency of electron  collisions with impurities
and phonons, $s = \sqrt{4\pi\,e^2\Sigma_0W/m\ae}$ is the plasma wave velocity,
and
$m$ is the electron effective mass. 
The effect of electron pressure results in some renormalization
of the plasma wave velocity:~\cite{23,24,25}  
$s \sqrt{(4\pi\,e^2\Sigma_0W/m\ae) + s_0^2}$,
where $s_0$ is the velocity of the electron ``sound'' 
(usually $s_0 \ll s$).

Assuming that the central gate length is comparable with the gate layer
thickness, the shape of the barrier
under the central gate can be considered as parabolic:
$\Delta_b(x) = \Delta_b[1 - (x/l)^2] - e(V_d + \delta V^{b})/2$,
where $\Delta_b$ is the barrier height in equilibrium
(it depends on  potential, $V_{cg}$, of the central  gate),
and $V_d + \delta V^{b}$ is the lateral potential drop
across the depletion (barrier) region associated with the dc bias 
voltage $V_d$ and the ac voltage $\delta V^{b}$.
The variation of the barrier 
height  $\delta \Delta_b$ due to  the lateral potential drop
is given by 
\begin{equation}\label{eq5}
\delta \Delta_b = - \frac{e(V_d + \delta V^{b})}{2}\biggl[1 - 
\frac{e(V_d + \delta V^{b})}{8\Delta_b}\biggr] \simeq
- \frac{e(V_d + \delta V^{b})}{2}.
\end{equation}
Hence, at $e|V_d + \delta V^{b}| \ll \Delta_b$
the net
source-drain current can be calculated using the following formula:
\begin{equation}\label{eq6}
J = J_m \exp\biggl(- \frac{\Delta_b}{k_BT}\biggr)
\biggl\{\exp\biggl[ \frac{e(V_d + \delta V^{b})}{2k_BT}\biggr] - 1\biggr\},
\end{equation}
where $J_m$ is the maximum  current (for a 2DEG $J_m \propto T^{3/2}$)
 and 
the ac potential drop, $\delta V^{b}$,  
across the depletion (barrier) region is given by

\begin{equation}\label{eq7}
\delta V^{b} = \delta\varphi\,|_{x = + l}- \delta\varphi\,|_{x = - l}
= 2 \delta\varphi\,|_{x = + l},
\end{equation}
and $T = T_0 + \delta T$ is the electron temperature,
$T_0$ is the lattice temperature, and $\delta T$ is the variation
of the electron temperature due to THz irradiation.

In this case, at $e\delta V^{b} < k_BT < \Delta_b
$, Eq.~(6) yields the following
equation for the dc source-drain current $J_0$:
\begin{eqnarray}\label{eq8}
  \Delta J_0 & =& J_0 - J_{00} = J_m\exp\biggl(- \frac{\Delta_b}{k_BT_0}\biggr)
  \exp\biggl( \frac{eV_d}{2k_BT_0}\biggr)
  \nonumber \\
  & &
  \times\biggl[\frac{e^2\overline{(\delta V^{b})^2} 
    + 8(\Delta_b - eV_d/2)k_B\overline{\delta T}}
  {8(k_BT_0)^2}\biggr]
\end{eqnarray}
at $eV_d > k_BT_0$ and
\begin{eqnarray}\label{eq9}
  \Delta J_0 &=& J_0 - J_{00} = J_m\exp\biggl(- \frac{\Delta_b}{k_BT_0}\biggr)
  \biggl( \frac{eV_d}{2k_BT_0}\biggr)
  \nonumber \\
  & &  \times
  \biggl[\frac{e^2\overline{(\delta V^{b})^2} 
    + 8(\Delta_b - eV_d/2)k_B\overline{\delta T}}
  {8(k_BT_0)^2}\biggr]
\end{eqnarray}
at $eV_d \ll k_BT_0$.
Here,
\begin{equation}\label{eq10}
J_{00}= J_m\exp\biggl(- \frac{\Delta_b}{k_BT_0}\biggr)
\biggl[\exp\biggl(\frac{eV_d}{2k_BT_0}\biggr) - 1\biggr]
\end{equation}
is the dc source-drain current 
without THz irradiation (dark current).

\section*{III.~PLASMA RESONANCES}

Solving Eq.~(4) with boundary conditions (2) and (3) and taking into account
Eq.~(7),
we obtain
\begin{equation}\label{eq11}
\delta V^b = \frac{V_{\omega}}{2}\biggl\{\frac{e^{i\omega\,t}}
{\cos [q_{\omega}(L - l)]} + 
\frac{e^{- i\omega\, t}}{\cos [q_{-\omega}(L - l)]}\biggr\}.
\end{equation}
Here 
\begin{equation}\label{eq12}
 q_{\pm\omega} = \sqrt{\frac{m\ae\omega(\omega \mp i\nu)}{4\pi e^2\Sigma_0W}}.
\end{equation}
Equation~(11) yields
\begin{equation}\label{eq13}
\overline{ (\delta V^b)^2} = \frac{V_{\omega}^2}{2}
\biggl[\cos^2\biggl(\frac{\pi\omega}{2\Omega}\biggr) + 
\sinh^2\biggl(\frac{\pi\nu}{4\Omega}\biggr)\biggr]^{-1},
\end{equation}
where
\begin{equation}\label{eq14}
\Omega = \sqrt{\frac{\pi^3e^2\Sigma_0W}{\ae\,m(L - l)^2}}
\end{equation}
is the characteristic plasma frequency.
Due to the dependence (given by Eq.~(1)) of 
the dc electron density on the gate
voltage $V_g$,  the characteristic plasma frequency 
$\Omega$ can be tuned by this voltage.
Because of the dependence of 
$\Omega$ on the length, $L - l$ of the quasi-neutral
sections of the getated 2DEG  channel
(serving as the resonant plasma cavities) and, hence, on
the potential
of the central gate $V_{cg}$,
$\Omega$
is somewhat varied with varying $V_{cg}$.

\section*{IV.~RESONANT ELECTRON HEATING}

The power absorbed by the channel 
can be calculated using the following equation:
\begin{equation}\label{eq15}
\delta P_{\omega} = \frac{2e^2\Sigma_0\,\nu}{m(\omega^2 + \nu^2)}
\int_l^{L}d\,x\,
\biggl|\frac{\partial\delta \varphi}{\partial\,x}\biggr|^2.
\end{equation}
Considering Eq.~(15), the energy balance equation governing the averaged
electron temperature can be presented as
\begin{equation}\label{eq16}
\frac{k_B\overline{\delta T}}{\tau_{\varepsilon}}
= \frac{e^2 \nu}{m(\nu^2 + \omega^2)}
\overline{\biggl(\frac{\delta V^{b}}{L - l}\biggr)^2}\,K,
\end{equation}
where $\tau_{\varepsilon}$ is the electron energy relaxation time.
The factor $K$ is associated with the nonuniformity of
the ac electric field under the side gates. This factor is given by
\begin{eqnarray}\label{eq17}
  K &=& \biggl(\frac{\pi}{8}\biggr)^2 
  \frac{\omega\sqrt{\omega^2 + \nu^2}}{\Omega^2}
  \biggl[\biggl(\frac{2\Omega}{\pi\nu}\biggr)\,\sinh
  \biggl(\frac{\pi\nu}{2\Omega}\biggr)
  \nonumber \\
  & &
  - \biggl(\frac{\Omega}{\pi \omega}\biggr)\,
  \sin
  \biggl(\frac{\pi\omega}{\Omega}\biggr)
  \biggr].
\end{eqnarray}
At $\nu \ll \omega, \Omega$, Eq.~(17) yields 
\begin{equation}\label{eq18}
K = \biggl(\frac{\pi}{8}\biggr)^2 \biggl(\frac{\omega}{\Omega}\biggr)^2
\biggl[1 - \biggl(\frac{\Omega}{\pi \omega}\biggr)\,
\sin\biggl(\frac{\pi\omega}{\Omega}\biggr)\biggr],
\end{equation}
so that one can use the following simplified formula:

\begin{equation}\label{eq19}
k_B\overline{\delta T} \simeq \biggl(\frac{\pi}{8}\biggr)^2
\frac{e^2 \nu\tau_{\varepsilon}}{mL^2\Omega^2}
\overline{(\delta V^{b})^2}.
\end{equation}

\section*{V.~RECTIFIED CURRENT AND DETECTOR RESPONSIVITY}

Using Eqs.~(8) and (15), we obtain

\begin{equation}\label{eq20}
\Delta J_0 = J_m\exp\biggl( \frac{V_d - 2\Delta_b}{2k_BT_0}\biggr)
\cdot\frac{(1 + H)e^2\overline{(\delta V^{b})^2}}
{8(k_BT_0)^2},
\end{equation}
where the term
\begin{equation}\label{eq21}
H =  \frac{4\,(2\Delta_b - eV_d)\nu\tau_{\varepsilon}}
{m(\nu^2 + \omega^2)(L - l)^2}\,K 
\end{equation}
is associated with the contribution of the electron heating to the
rectified  current.

Combining Eqs.~(13) and (20),
we arrive at
\begin{equation}\label{eq22}
\Delta J_0 =J_m
\cdot\displaystyle\frac{\exp\biggl( 
\displaystyle \frac{V_d - 2\Delta_b}{2k_BT_0}\biggr)\,(1 + H)}
{4\displaystyle\biggl[\cos^2\biggl(\frac{\pi\omega}{2\Omega}\biggr) + 
\sinh^2\biggl(\frac{\pi\nu}{4\Omega}\biggr)\biggr]}
\biggl(\frac{eV_{\omega}}{2k_BT_0}\biggr)^2.
\end{equation}
Since $V_{\omega}^2$ is proportional to the incoming THz power,
the detector responsivity as a function of the signal frequency
and the structural parameters (except the antenna parameters) can be
in the following form
\begin{equation}\label{eq23}
R \propto 
\displaystyle\frac{\exp\biggl( 
\displaystyle \frac{V_d - 2\Delta_b}{2k_BT_0}\biggr)}
{\displaystyle\biggl[\cos^2\biggl(\frac{\pi\omega}{2\Omega}\biggr) + 
\sinh^2\biggl(\frac{\pi\nu}{4\Omega}\biggr)\biggr]}
\frac{(1 + H)}{\sqrt{k_BT_0}}.
\end{equation}
As follows from Eq.~(23), the responsivity as a function
of the frequency of incoming THz radiation
exhibits sharp
peaks at the plasma resonant frequencies $\omega = \Omega(2n - 1)$,
where $n = 1, 2, 3, ...$ is the resonance index, provided that the quality
factor of resonances $\Omega/\nu \gg 1$.

\section*{VI.~COMPARISON OF DYNAMIC AND HEATING MECHANISMS}

As seen from Eq.~(22), 
the ratio of the   heating and  dynamic components
of the rectified dc source-drain current
is given by the quantity $H$ (heating parameter).
At $\nu \ll \omega, \Omega$,
considering Eqs.~(18) and (19),
the quantity $H$ can be estimated as
\begin{equation}\label{eq24}
H =  \frac{\pi^2}{16}\frac{(2\Delta_b - eV_b)\nu\tau_{\varepsilon}}
{m\Omega^2(L - l)^2}. 
\end{equation}
Considering that 
$\Omega = \pi s/2(L - l)$, Eq.~(24) can be presented in the following form:
\begin{equation}\label{eq25}
H \simeq \frac{1}{4}\biggl(\frac{2\Delta_b - eV_g}{ms^2}\biggr)
\nu\tau_{\varepsilon}.
\end{equation}

The first factor in Eq.~(25) is usually small, 
while the second one can be large,
so that $H$ can markedly exceed unity,
 particularly, at low temperatures.

It is natural to assume that 
in a wide range of the temperature 
(from liquid helium to room temperatures)
the electron collision frequency (the inverse momentum relaxation time)
is determined by the electron interaction with acoustic phonons
and charged impurities, while the electron energy relaxation
is due to the interaction with both acoustic and polar optical phonons.
Hence,
$\nu = \nu^{(ac)} + \nu^{(i)}$ and 
$\tau_{\varepsilon} = 
\tau_{\varepsilon}^{(ac)}\tau_{\varepsilon}^{(op)}
/[\tau_{\varepsilon}^{(ac)} + \tau_{\varepsilon}^{(op)}]$,
where $\nu^{(ac)}$ and $\nu^{(i)}$ are related to the electron scattering
on acoustic phonons  and impurities (charged), respectively,
whereas $\tau_{\varepsilon}^{(ac)}$ and $\tau_{\varepsilon}^{(op)}$
are the electron energy relaxation times associated with
the acoustic and optical phonons.
As a result,
 the product $\nu\tau_{\varepsilon}$ can be presented as
\begin{equation}\label{eq26}
\nu\tau_{\varepsilon}
 = \frac{[\nu^{(ac)} + \nu^{(i)}]
\tau_{\varepsilon}^{(ac)}\tau_{\varepsilon}^{(op)}}{[\tau_{\varepsilon}^{(ac)} 
+ \tau_{\varepsilon}^{(op)}]}.
\end{equation}

We shall use the following formulas for the temperature dependences
of the parameters in this equation:
\begin{equation}\label{eq27}
\nu^{(ac)} = \frac{1}{\overline{\tau}^{(ac)}_{PA}}
\biggl(\frac{k_BT_0}{\hbar\omega_0}\biggr)^{1/2} +
 \frac{1}{\overline{\tau}^{(ac)}_{DA}}
\biggl(\frac{k_BT_0}{\hbar\omega_0}\biggr)^{3/2},
\end{equation}
\begin{equation}\label{eq28}
\frac{1}{\tau_{\varepsilon}^{(ac)}} = \frac{2ms_a^2}{\hbar\omega_0}
\biggl[\frac{1}{\overline{\tau}^{(ac)}_{PA}}
\biggl(\frac{k_BT_0}{\hbar\omega_0}\biggr)^{- 1/2} +
 \frac{1}{\overline{\tau}^{(ac)}_{DA}}
\biggl(\frac{k_BT_0}{\hbar\omega_0}\biggr)^{1/2}\biggr],
\end{equation}
\begin{equation}\label{eq29}
\frac{1}{\tau_{\varepsilon}^{(op)}} = \frac{1}{\overline{\tau}^{(op)}}
\frac{2}{3}\biggl(\frac{\hbar\omega_0}{k_BT_0}\biggr)^2
\exp\biggl(- \frac{\hbar\omega_0}{k_BT_0}\biggr),
\end{equation}
where $\overline{\tau}^{(ac)}_{PA} = 8$~ps, 
$\overline{\tau}^{(ac)}_{DA} = 4$~ps,
and $\overline{\tau}^{(op)} = 0.14$~ps~\cite{26} 
are the characteristic scattering times
for the acoustic phonon scattering
(polar and deformation mechanisms, respectively)
and for the optical phonon scattering, and $\hbar\omega_0$ is the optical
phonon energy.
The temperature
dependence of the collision frequency of 
2D electrons with charged impurities
is assumed to be as follows:
\begin{equation}\label{eq30}
\nu^{(i)} = \overline{\nu}^{(i)} \biggl(\frac{\hbar\omega_0}{k_BT_0}\biggr)\cdot
S(|z_i|, k_BT_0/\hbar\omega_0).
\end{equation}
Here the function~\cite{27} 
$$
S(z_i, \xi) = \int_0^1\exp (-\alpha|z_i| \sqrt{\xi}\,x)\frac{dx}{\sqrt{1 - x^2}}
$$
characterizes a decrease in the collision frequency with increasing
thickness of the spacer, $|z_i|$, between the 2DEG channel
and the charged donor sheet, and 
$\alpha = 4\sqrt{2m\hbar\omega_0}/\hbar \simeq 10^7$~cm$^{-1}$.
The characteristic frequency  $\overline{\nu}_i$
is proportional to the donor sheet density. We set  
$\overline{\nu}^{(i)} = (10^{8} - 10^{9})$~s$^{-1}$.
At $T_0 = 4.2$~K, these values correspond 
to $\nu^{(i)} = 10^{10} - 10^{11}$s$^{-1}$
and the electron mobilities $\mu \simeq (10^5 - 10^6)$~cm$^2$/V$\cdot$s.
When $|z_i| = 0$, Eq.~(30) yields $\nu_i \propto T_0^{-1}$
while at sufficiently thick spacers, one obtains 
$\nu_i \propto T_0^{-3/2}$. 
In the case of relatively thick
electron channels, in which the quantization is insignificant,
$\nu_i \propto T_0^{-3/2}$.

\begin{figure}[t]
  \begin{center}
    \includegraphics[width=8.5cm]{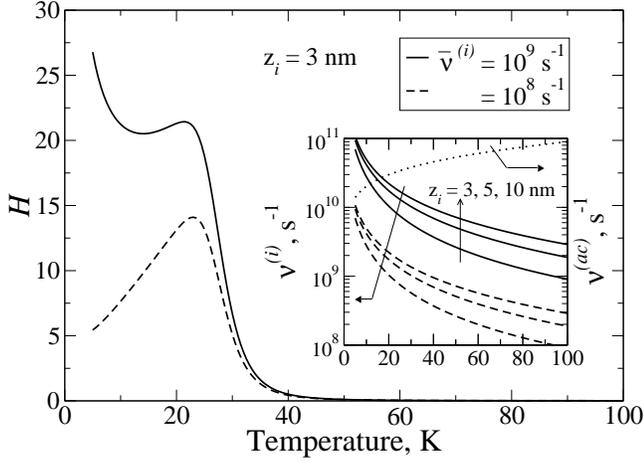}
    \caption{Temperature dependences of heating parameter $H$. The inset
      shows the temperature dependences of the electron collision frequencies
      associated with impurity and acoustic scattering mechanism.}
  \end{center}
\end{figure}

Figure~2 shows the temperature dependence of the ``heating''
parameter $H$, which determines the relative contribution
of the heating mechanism to the detector responsivity,
calculated 
for different values of parameter $\overline{\nu}^{(i)}$, 
i.e., different doping levels.
It is assumed that for a GaAs channel 
$\hbar\omega_0/k_B = 421$~K.~\cite{26} We set 
$s = 1\times 10^8$~cm/s, $s_a = 5\times10^5$~cm/s,
the barrier height $\Delta_b = 110$~meV,
and the bias voltage $V_d = 20$~mV, so that
the effective barrier height $\Delta_b^{(eff)}
= \Delta_b - eV_b/2 = 100$~meV.
As seen from Fig.~2, the heating mechanism can provide significantly
larger contribution to the detector responsivity at
low temperatures ($T_0 \lesssim 35 - 40$~K). 
However at elevated temperatures,
this mechanism becomes relatively inefficient.
This is attributed to weak electron heating 
due to strong energy relaxation
on optical phonons at elevated temperature. 



\section*{VII.~TEMPERATURE DEPENDENCES OF THE DETECTOR RESPONSIVITY
AND DETECTIVITY}
 
As mentioned in Sec.~V, the detector resonsivity exhibits 
sharp
peaks at the plasma resonant frequencies $\omega = \Omega(2n - 1)$.
The sharpness of these peaks depends on the electron collision frequency
$\nu$, which in turn depends on the temperature.

As follow from Eq.~(23),
the maximum values of the detector responsivity
at the fundamental resonance ($n = 1$) is given by 
\begin{equation}\label{eq31}
{\rm max} R \propto 
\displaystyle\frac{\exp\biggl( 
\displaystyle \frac{V_d - 2\Delta_b}{2k_BT_0}\biggr)}
{\displaystyle
\sinh^2\biggl(\frac{\pi\nu}{4\Omega}\biggr)}
\frac{(1 + H)}{\sqrt{k_BT_0}}.
\end{equation}
As a result,
for the value of the ratio of the detector responsivity 
at the resonance 
and the dark current one obtains
\begin{equation}\label{eq32}
\frac{{\rm max} R}{J_{00}} \propto 
\displaystyle\frac{1}
{\displaystyle
\sinh^2\biggl(\frac{\pi\nu}{4\Omega}\biggr)}\frac{1 + H}{(k_BT)^2}.
\end{equation}

Figure~3 shows the temperature dependences of 
 the maximum (resonant) values of the dynamic and heating
contributions to the detector responsivity devided by the dark current
value
calculated using Eq.~(32) for 
 the barrier height $\Delta_b^{(eff)} = 100$~meV, 
the fundamental plasma frequencies
$\Omega/2\pi = 1$~THz.
and different values  of the
parameter $\overline{\nu}^{(i)}$.
The inset shows the temperature dependence of
max $R/J_{00}$, which accounts for both mechanisms. 

\begin{figure}[t]
  \begin{center}
    \includegraphics[width=8.5cm]{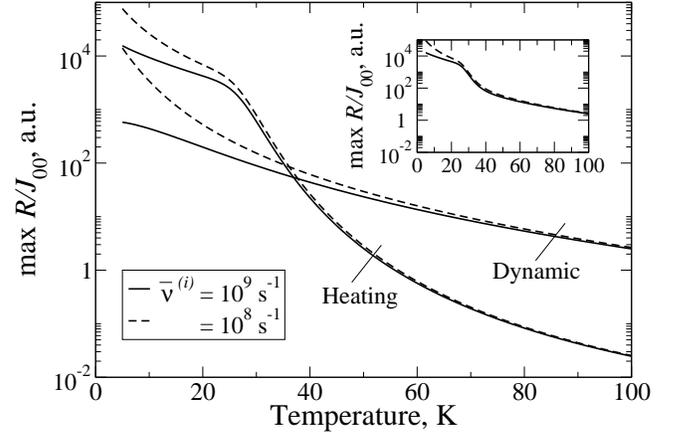}
    \caption{Temperature dependences of dynamic and heating contributions to 
      $max R/J_{00}$, of the detector responsivity maximum (resonant) 
      value and the dark current. The inset shows the net value
      $max R/J_{00}$ as a function of temperature.}
    \end{center}
\end{figure}

\begin{figure}[t]
  \begin{center}
    \includegraphics[width=8.5cm]{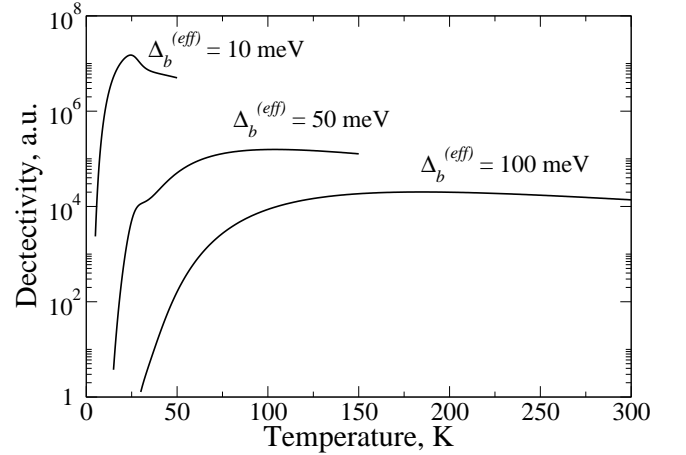}
    \caption{Temperature dependences of the detector responsivity
      $D^{*}$ for different values of the effective barrier height
      $\Delta_{b}^{(eff)}$.}
  \end{center}
\end{figure}

Taking into account that the detector detectivity
$D^* \propto R/\sqrt{J_{00}}$, we arrive at the following formula:

\begin{equation}\label{eq33}
{\rm max} D^* \propto 
\displaystyle\frac{\exp\biggl( 
\displaystyle \frac{V_d - 2\Delta_b}{4k_BT_0}\biggr)}
{\displaystyle\biggl[
\sinh^2\biggl(\frac{\pi\nu}{4\Omega}\biggr)\biggr]}
\frac{(1 + H)}{(k_BT_0)^{5/4}}.
\end{equation}

Figure~4 shows the temperature dependences of the detector detectivity
calculated for different values of the effective barrier height
$\Delta_b^{(eff)} = \Delta_b - eV_d/2$.

As follows from the above formulas,
the detector responsivity and detectivity increase
with decreasing effective barrier height.
However, this quantity can not be set too small (at a given temperature
$T_0$)
in the frame of our model, 
which is  valid if $\Delta_b^{(eff)} \gg k_BT_0$,
and when the 2DEG channel can be partitioned
into the quasi-neutral and deleted sections.


\section*{VIII.~DISCUSSION}

The displacement current across the depleted region can be,
in principle, essential.
To take this current into consideration
one needs to modify Eq.~(3).~\cite{28}
Introducing the admittance, $Y_{\omega}$,  of the barrier region,
we can use the following boundary conditions
at the edge of the quasi-neutral sections of the 2DEG channel
(for the asymmetrical plasma modes for which 
$\delta\varphi_{\omega}|_{x = - l} = - \delta\varphi_{\omega}|_{x = l}$):

\begin{equation}\label{eq34}
- \sigma_{\omega}\frac{d\varphi_{\omega}}{d\,x}\biggr|_{x = \pm l}
= 2Y_{\omega}\delta\varphi_{\omega}|_{x = \pm l}.
\end{equation}
Here, 
$\sigma_{\omega} = -i[e^2\Sigma_0/m(\omega + i\nu)]$
is the ac conductivity of the quasi-neutral sections of the channel
accounting for the electron collisions and inertia.
Since the real part of the conductivity of the depleted
(barrier) region is definitely small in comparison with the
conductivity of the qusi-neutral sections, one can
take into account only the capacitive component of the current
across the depleted region. In this case, $Y_{\omega} = -i\omega C$,
where $C$ is the pertinent capacitance, which is determined by
the capacitances $C_{cg}$ and $C_{dr}$:
$C = C_{dr} + C_{gs}/2$ (see the equivalent circuit
in Fig.~1). This equivalent circuit accounts for
the inductance of the quasi-neutral sections ${\cal L}_c$,
due to the inertia of the electron transport along the channel,
the channel resistance due to the electron scattering, 
the resistance of the barier region $R_{dr}$, and different capacitances.
Equation~(34) can be presented in the form:

\begin{equation}\label{eq35}
\frac{\Omega^2 (L - l)}{ \omega(\omega + i\nu)}\frac{d\delta \varphi_{\omega}}
{d\,x}\biggr|_{x = \pm l} = c\,\delta\varphi_{\omega}|_{x = \pm l},
\end{equation}
where
\begin{equation}\label{eq36}
c = \frac{2\pi^2 CW}{\ae\,(L - l)} \simeq 
\frac{2\pi^2 CW}{\ae\,L}.
\end{equation}
If the parameter $c \ll 1$,
Eqs.~(3) and (35) practically coincide 
(except the case  $\omega \gg \Omega$).
According to Ref.~\cite{28} 
$C_{dr} = (\ae/2\pi^2)\Lambda_{dr}$ and 
$C_{cg} = (\ae/2\pi^2)\Lambda_{cg}$,
where  $\Lambda_{dr}$ and $\Lambda_{dr}$ are logarithmic
factors which are deternined by the geometry
of the planar conducting areas (the gates and quasi-neutral sections of
the channel). These factors can be estimated as:~\cite{28} 
 $\Lambda_{dr} \sim \Lambda_{cg} \sim \ln (2L/l)$.
Hence, $c \simeq (W/L)\Lambda$, where
$\Lambda = \Lambda_{dr} + \Lambda_{cg}/2$.
The factor $\Lambda$ somewhat exceeds unity although it is not too large.
Thus, taking into account that $W \ll L$,
one can conclude that for real device structures $c \ll 1$. 
The condition of the smallness of $c$ can
be also presented in the form $C_g \gg C$,
where $C_g = \ae\/4\pi\,W  \simeq L/W$ is the capacitance
of the side gates (see Fig.~1).
Relatively small capacitances $C_{dr}$ and $C_{cg}$
can, to some extent, affect the plasma oscillations.
However, their role reduces mainly to a small modification
in the resonant plasma frequencies.~\cite{28}

If $c$ would be large, the ac potential near the edges
of the quasi-neutral regions is small as well as the ac potential
drop across the depleted
region. This implies that the boundary condition~(34) 
could be
$\delta\varphi|_{\omega}|_{x = \pm l} \simeq 0$.
In such a case, the fundamental plasma frequency
is doubled and dynamic mechanism is weakened. However,
the details of the ac potential distribution near the edges
of the quasi-neutral regions are not crucial for the heating
mechanism.

Actually, there is some delay in the electron transit
across the barrier. The delay time can be estimated as
$\tau_b \simeq 2l/v_T$, where $v_T \simeq \sqrt{k_BT_0/m}$
is the thermal electron velocity.
The delay in the electron transit and, therefore, 
the electron transit time effects can be disregarded if $\omega \tau_b
\sim \Omega\tau_b < 1$.~\cite{29} Assuming $\Omega/2\pi = 1$~THz and
$v_T = 10^7$~cm/s, from the last inequality we obtain
$2l < 0.6~\mu$m.

\section*{IX.~CONCLUSIONS}

We developed a device model for a resonant detector of THz radiation
based on the gated 2DEG channel with an electrically induced barrier.
The model accounts for the resonant excitation
of the plasma oscillations in the gated 2DEG channel
as well as the rectified dc current through the barrier.
As shown,  this current comprises two
components: the dynamic component associated with  
the ac potential drop across the barrier 
and the heating (bolometric) component due to 
a change in the electron temperature which  stems from the resonant
plasma-assisted absorption of THz radiation.
Using our model, we calculated the frequency and temperature
dependences of the detector responsivity. 
The detector responsivity exhibits sharp resonant peaks
at the frequencies of incoming THz radiation
corresponding to  the plasma  resonances in the gated 2DEG
channel. The plasma resonances can tuned by
the gate voltage $V_{cg}$ and, to some extent, by the
potential of the central gate $V_{cg}$. 
It is demonstrated
that the dynamic mechanism dominates at elevated
temperatures, whereas at low lattice temperatures 
($T_0 \lesssim 35 - 40$~K 
for AlGaAs/GaAs based detectors), the heating mechanism prevails. 
This can be attributed to a marked increase in the electron
energy relaxation time $\tau_{\varepsilon}$ with decreasing
lattice temperature.

At rather low temperatures when the ratio $\Delta_b^{(eff)}/k_BT_0$
is large, the thermionic electron current over the barrier
can be surpassed by the tunneling current.
The rectified portion of this current can be associated with
both dynamic and 
heating  mechanisms (the thermo-assisted tunneling current
in the case of the latter mechanism). This, however, requires
a separate detailed study.  

In the case of detection of  THz radiation modulated at some
frequency $\omega_m \ll \omega$, the relative contributions
of the dynamic and heating mechanisms
to the  responsivity, $R^{m}$, characterising
the detector response at the frequency $\omega_m$
can be different than those considered above.
This is due to different inertia of these mechanisms.
In particular, if $\omega_m > \tau_{\varepsilon}^{-1}$
(but $\omega_m \ll \omega$),
the variation of the electron effective temperature
averaged over the THz oscillations 
$\overline{\delta T_m} \simeq \overline{\delta T}/
\omega_m\tau_{\varepsilon}$,
i.e., $\overline{\delta T_m} \ll \overline{\delta T}$.
As a result, $R_m/R \simeq (\omega_m\tau_{\varepsilon})^{-1}$.
This implies that even at low temperatures,
the heating mechanism can be inefficient if 
$\omega_m > \tau_{\varepsilon}^{-1}$.

Both dynamic and heating mechanisms might be responsible for
the THz detection in the plasmonic resonant detectors utilizing
another barrier structure (for instance, with the lateral 
Schottky barrier~\cite{19,20} or with electron transport
through the gate barrier~\cite{30,31,32}) and  another
types of the plasma resonant cavity (with ungated quasi-neutral
regions as in high-electron mobility transistors with relatively
long ungated source-gate and gate-drain regions), as well as
in those with  another methods of the excitation
of plasma oscillations (utilizing periodic gate structures~\cite{14}).


\section*{Acknowledgments}

 This work was 
 supported by the  Grant-in-Aid for Scientific Research (S)
from the Japan Society for Promotion of Science, Japan.
The work at RPI was partially supported by the Office of Naval
Research, USA.

\end{document}